\documentclass[runningheads]{llncs}
\usepackage[T1]{fontenc}
\usepackage[table]{xcolor}
\usepackage{pifont}
\newcommand{\cmark}{\ding{51}}
\newcommand{\xmark}{\ding{55}}
\newcommand{\pmark}{$\circ$}
\usepackage{multirow}

\usepackage{graphicx}
\usepackage{cite}
\usepackage{amsmath,amssymb,amsfonts}
\usepackage{algorithmic}
\usepackage{graphicx}
\usepackage{textcomp}
\usepackage{xcolor}
\usepackage{comment}
\usepackage{booktabs}
\usepackage{subfig}
\usepackage{array}
\usepackage{tabularx}

\newcolumntype{Y}{>{\centering\arraybackslash}X}
\begin{document}
\title{Towards Fine-Grained Multi-Dimensional Speech Understanding: Data Pipeline, Benchmark, and Model}
%
\titlerunning{Towards Fine-Grained Multi-Dimensional Speech Understanding}
%
\author{Guojian Li$^{1}$, Zhixian Zhao$^{1}$, Zhennan Lin$^{1}$,  Jingbin Hu$^{1}$, Qirui Zhan$^{1}$, \\ Yuang Cao$^{1}$,
    Pengyuan Xie$^{2}$, Chuan Xie$^{2}$, Jie Liu$^{2}$, Qiang Zhang$^{2}$, \\ Zhonghua Fu$^1$$~^*$, Lei Xie$^1$$~^*$}
\authorrunning{Guojian Li et al.}
%
\institute{$^{1}$Audio, Speech and Language Processing Group (ASLP@NPU), School of Computer Science, Northwestern Polytechnical University, Xi’an, China \\
    $^{2}$Shanghai Lingguang Zhaxian Technology \\
\email{aslp\_lgj@mail.nwpu.edu.cn, mailfzh@nwpu.edu.cn, lxie@nwpu.edu.cn}}
\maketitle              
\begin{abstract}
While speech Large Language Models (LLMs) excel at conventional tasks like basic speech recognition, they lack fine-grained, multi-dimensional perception. This deficiency is evident in their struggle to disentangle complex features like micro-acoustic cues, acoustic scenes, and paralinguistic signals. This resulting incomplete comprehension of real-world speech fundamentally bottlenecks the development of perceptive and empathetic next-generation speech systems. At its core, this persistent perceptual limitation primarily stems from three interacting factors: scarce high-quality expressive data, absent fine-grained modeling for multi-dimensional attributes, and reliance on restricted coverage, coarse-grained benchmarks. We address these challenges through three pillars: First, our robust data curation pipeline resolves complex acoustic environments and long-audio timestamp alignment challenges to extract a high-quality spontaneous speech corpus from audiovisual sources. Second, we construct \emph{FMSU-Bench}, a pioneering benchmark covering 14 speech attribute dimensions to rigorously assess the fine-grained, multi-dimensional speech understanding capabilities of current models. Third, empowered by our curated corpus, we introduce \emph{FM-Speech}. Driven by a decoupled attribute modeling and progressive curriculum fine-tuning framework, it substantially elevates fine-grained, multi-dimensional acoustic perception. Extensive evaluations on \emph{FMSU-Bench} reveal that current speech understanding models still require significant improvement in multi-dimensional, fine-grained understanding. In contrast, \emph{FM-Speech} substantially outperforms current open-source models, establishing a robust paradigm for real-world speech understanding. Code, benchmark, and model will be open-sourced on GitHub\footnote{https://github.com/ASLP-lab/FMSU}.

\end{abstract}

\begingroup
\renewcommand\thefootnote{$^{*}$}
\footnotetext{Corresponding author.}
\endgroup

\section{Introduction}
Although speech LLMs have advanced speech understanding, they remain confined to conventional tasks such as speech recognition and lack fine-grained, multi-dimensional perception. This deficiency manifests in the difficulty of these models to disentangle complex features, such as micro-acoustic cues, acoustic scenes, and paralinguistic signals. The resulting incomplete comprehension of real-world speech hinders the development of perceptive and empathetic next-generation speech artificial intelligence. Fundamentally, this limitation stems from three primary factors: the scarcity of high-quality expressive data, the absence of fine-grained modeling for multi-dimensional attributes, and the reliance on benchmarks with limited coverage and coarse granularity. We address these challenges through three pillars: data curation, benchmarking, and modeling paradigms.

From the data perspective, widely used corpora fall short in two ways: controlled read speech (e.g., LibriSpeech~\cite{panayotov2015librispeech}) inherently lacks expressiveness, whereas expressive datasets (e.g., Emilia~\cite{he2024emilia}) lack unified, fine-grained, multi-dimensional annotations. To address this gap, we develop a robust, LLM-driven curation pipeline augmented by multi-expert cross-verification. Specifically tailored for highly expressive in-the-wild audiovisuals (e.g., movies, TV shows), this pipeline effectively resolves severe challenges in complex acoustics and long-audio timestamp alignment, thereby generating a high-quality speech corpus with fine-grained, multi-dimensional annotations.

From the evaluation perspective, existing speech understanding benchmarks (e.g., AIR-bench~\cite{yang2024air}, MMAR \cite{ma2025mmar}, MMAU~\cite{sakshi2024mmau}) suffer from limited dimensional coverage and coarse annotation granularity. Regarding dimensional coverage, they predominantly cater to macroscopic tasks like basic speech recognition and broad emotion categorization, or exhibit a strong bias toward semantic-dependent reasoning tasks. Regarding annotation granularity, these benchmarks evaluate models by requiring monolithic, coarse-grained labels. This paradigm fails to assess fine-grained perceptual acuity, frequently overlooking subtle yet critical micro-acoustic features essential for real-world speech comprehension. To overcome these evaluation deficiencies, we construct \textbf{FMSU-Bench}, a pioneering \textbf{F}ine-grained \textbf{M}ulti-dimensional \textbf{S}peech \textbf{U}nderstanding \textbf{Benchmark}. Comprising over 20,000 bilingual instances, FMSU-Bench systematically covers 14 distinct speech dimensions structured into a comprehensive 5-tier taxonomy: \textbf{Speaker Demographics}, \textbf{Acoustic-Prosodic Features}, \textbf{Affective and Semantic Reasoning}, \textbf{Acoustic Scene Analysis}, \textbf{Linguistic-Paralinguistic Integration}. This hierarchical design enables a holistic assessment of how well a model comprehends fine-grained, multi-dimensional real-world speech.

From the modeling perspective, current speech understanding LLMs (e.g., Qwen3-Omni~\cite{xu2025qwen3omni}, Audio Flamingo 3~\cite{goel2025audioflamingo3}) face several critical limitations. They often exhibit restricted or entangled modeling of speech attributes, frequently yielding coarse or single-label outputs. Furthermore, they suffer from hallucinations driven by text, overly depending on linguistic priors while neglecting actual acoustic evidence. To overcome these limitations, we introduce \textbf{FM-Speech}, a \textbf{F}ine-grained \textbf{M}ulti-dimensional \textbf{Speech} understanding model. Leveraging the multi-dimensional fine-grained annotations produced by our data pipeline, FM-Speech employs a progressive curriculum fine-tuning framework designed to achieve deep, decoupled modeling of complex speech attributes. It jointly captures 14 distinct speech attributes within a unified paradigm, significantly enhancing fine-grained perception and outperforming existing open-source speech LLMs on our FMSU-Bench.

Our core contributions are threefold: \textbf{1) Data Pipeline:} We develop a robust, LLM-driven data curation pipeline augmented by multi-expert cross-verification. Resolving complex acoustic environments and timestamp alignment challenges, it extracts a high-quality speech corpus with fine-grained, multi-dimensional annotations from audiovisual sources. \textbf{2) Benchmark:} We release \textbf{FMSU-Bench}, a rigorous bilingual benchmark spanning 14 dimensions, establishing a new standard for evaluating fine-grained, multi-dimensional auditory perception. \textbf{3) Model:} We propose a progressive curriculum fine-tuning framework to train \textbf{FM-Speech}, which disentangles intertwined acoustic attributes, simultaneously outputs fine-grained annotations across multiple dimensions, and achieves state-of-the-art (SOTA) performance among open-source models on FMSU-Bench.

\section{METHODS}

\subsection{Data Curation Pipeline}
\begin{figure}[t]
    \centering
    \includegraphics[width=0.85\columnwidth]{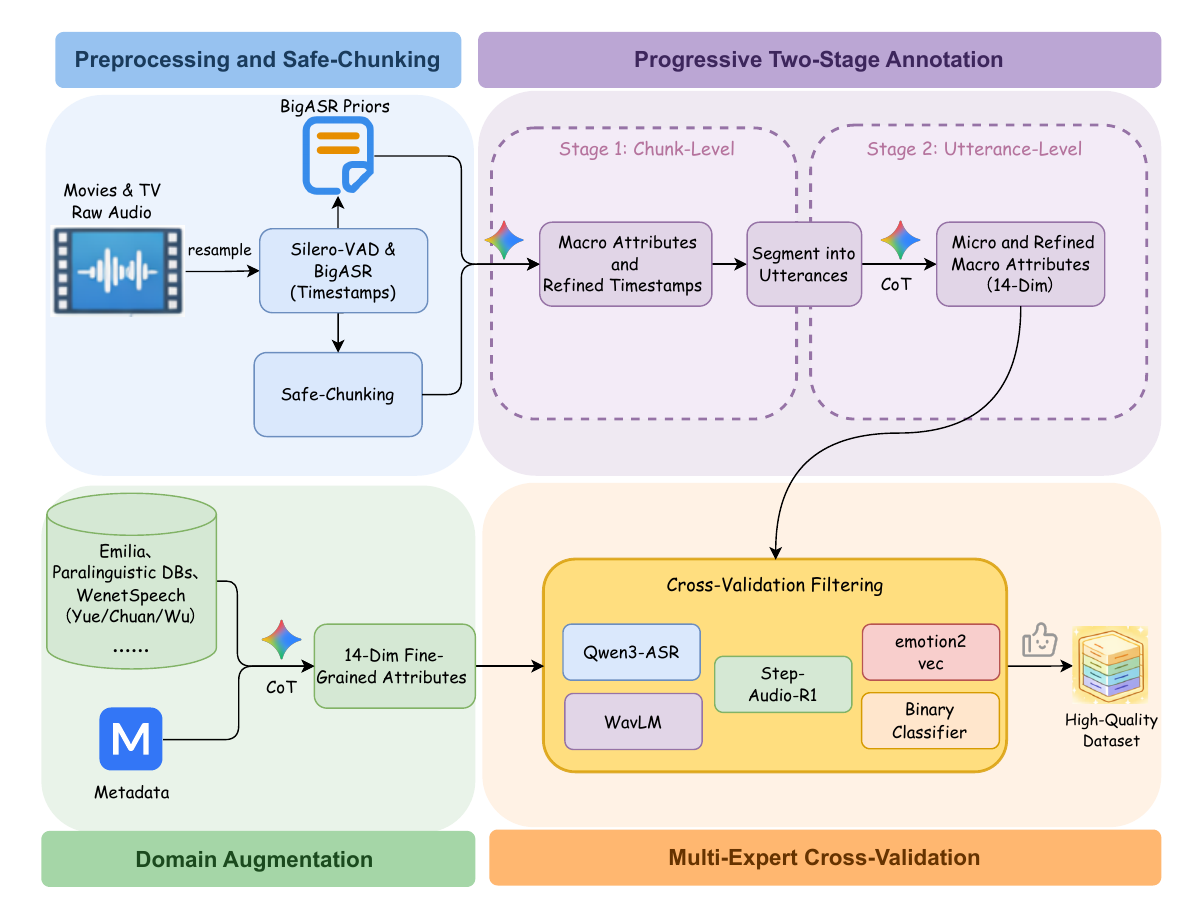}
    \caption{The overview of our proposed data curation pipeline.}
    \label{fig:overview}
\end{figure}

To capture highly expressive spontaneous speech, we primarily source raw data from movies and TV shows. As illustrated in Figure~\ref{fig:overview}, our LLM-driven data curation pipeline, augmented by multi-expert cross-verification, comprises four core components:

\subsubsection{Preprocessing and Safe-Chunking}
Initially, all raw audio from movies and TV shows is uniformly resampled to 16 kHz mono, 16-bit, yielding typically lengthy recordings. To facilitate processing, we segment these long audio into shorter chunks. This segmentation presents a critical trade-off: overly long chunks exacerbate LLM temporal hallucinations, while excessively short ones lose essential contextual cues, and arbitrary cuts risk mid-sentence truncation. To resolve this, we propose a safe-chunking strategy with an optimal 5–6 minute window, empirically chosen to balance temporal hallucination against context loss. We jointly utilize Silero-VAD \cite{Silero-VAD} and Volcengine BigASR\footnote{https://www.volcengine.com/docs/6561/1354869} to extract utterance-level timestamps. By merging their detected speech intervals, we robustly identify true silence regions and segment the audio exactly at their midpoints within the 5–6 minute window. This effectively prevents speech truncation, preserving both acoustic and semantic integrity for subsequent annotations.

\subsubsection{Progressive Two-Stage Annotation}
Leveraging Gemini 2.5 Pro \cite{comanici2025gemini}, we design a progressive ``macro-to-micro'' two-stage annotation strategy. Specifically, in the chunk-level first stage, we mitigate Gemini's timestamp hallucinations by feeding Volcengine BigASR's timestamps, transcriptions, and speaker IDs as strong priors alongside the 5–6 minute audio chunks. Constrained by these priors, the model calibrates timestamps, rectifies transcriptions, and annotates context-dependent macro attributes: \textit{contextual inference}, \textit{background sound}, \textit{acoustic environment}, and \textit{transcription with paralinguistic tags}. Subsequently, in the utterance-level second stage, we segment the chunks into individual utterances using the refined timestamps, utilizing the macro attributes derived in the first stage as contextual priors. Guided by Chain-of-Thought (CoT) prompts to reason from low-level acoustics to high-level traits, Gemini 2.5 Pro annotates the micro-attributes (\textit{gender}, \textit{age}, \textit{accent}, \textit{pitch}, \textit{speaking rate}, \textit{rhythm}, \textit{voice texture}, \textit{emotion}, \textit{tone}, and \textit{paralinguistic events}) while concurrently cross-validating and refining the inherited macro attributes. Following this cohesive process, each utterance yields a structured JSON file that intricately details all 14 distinct speech dimensions.

\subsubsection{Domain Augmentation}
To address attribute sparsity and enrich the dataset, we incorporate supplementary open-source datasets. We sample from Emilia \cite{he2024emilia} and several paralinguistic corpora (NonVerbalSpeech-38K \cite{ye2025scalable38k}, Emilia-NV \cite{liao2025nvspeech}, SMIIP-NV \cite{wu2025smiip}, and NonverbalTTS \cite{borisov2025nonverbaltts}) to augment paralinguistic instances. To enhance accent diversity, we extract utterances from WenetSpeech-Yue \cite{li2026wenetspeechyue}, WenetSpeech-Chuan \cite{dai2025wenetspeechchuan}, WenetSpeech-Wu \cite{wang2026wenetspeechwu}, and Common-Voice-English \cite{ardila2020commonvoice}. Since these are pre-segmented, we utilize their original metadata as priors. Gemini 2.5 Pro processes these via CoT prompting to generate structured JSON annotations identical to our established schema, ensuring corpus-wide consistency.

\subsubsection{Multi-Expert Cross-Validation}
To address Gemini 2.5 Pro's occasional judgment biases and hallucinations, we introduce a rigorous cross-validation mechanism using multiple expert models. Specifically, we first deploy Qwen3-ASR-1.7B~\cite{shi2026qwen3asr} to re-transcribe individual utterances, discarding any sample with a Word Error Rate (WER) or Character Error Rate (CER) exceeding 30\%. For affective and low-level acoustic features, we utilize emotion2vec-large~\cite{ma2024emotion2vec} to retain only samples whose predicted emotional polarity aligns with the Gemini 2.5 Pro annotations. We simultaneously utilize Step-Audio-R1~\cite{tian2025stepaudior1} to evaluate pitch and speaking rate, cross-referencing these acoustic metrics against Gemini 2.5 Pro outputs to retain instances with consistent intensity levels. Additionally, to ensure the reliability of speaker demographics, we apply WavLM-Large-based classifiers from VoxProfile~\cite{feng2025voxprofile} to filter samples based on the intersection of age, gender, and accent predictions. Finally, a Wav2Vec-BERT 2.0-based binary classifier~\cite{ye2025scalable38k} is deployed to meticulously eliminate conflicting speech instances where the binary model and Gemini 2.5 Pro disagree on the presence of paralinguistic events.

\begin{table*}[htbp]
\centering
\caption{Benchmark overview—task abbreviations, examples, and item counts.}
\label{tab:fmsu_bench_tasks}
\begin{tabularx}{\textwidth}{@{} p{0.1\textwidth} X >{\centering\arraybackslash}p{0.13\textwidth} @{}}
\toprule
\textbf{Task} & \multicolumn{1}{>{\centering\arraybackslash}X}{\textbf{Example}} & \textbf{ZH / EN} \\ \midrule

\multicolumn{3}{c}{\textbf{Speaker Demographics}} \\ \midrule
GEN & \textbf{Q:} What is the gender of the speaker in this speech? \textbf{A.} Male \textbf{B.} Female... & 932 / 911 \\
AGE & \textbf{Q:} Based on the speaker's voice, what is their perceived age group? \textbf{A.} Young Adult (19-35 years old) \textbf{B.} Teenager (13-18 years old)... & 787 / 728 \\
ACC & \textbf{Q:} Which of the following best describes the speaker's accent? \textbf{A.} The speaker has a North American accent. \textbf{B.} The speaker has an English accent... & 1000 / 900 \\ \midrule

\multicolumn{3}{c}{\textbf{Acoustic-Prosodic Features}} \\ \midrule
PIT & \textbf{Q:} Please determine the pitch characteristics of this speech. \textbf{A.} Low-pitched with a relatively flat intonation. \textbf{B.} Low-pitched with severe fluctuations... & 853 / 958 \\
SR & \textbf{Q:} What are the speaking rate characteristics of this speech? \textbf{A.} Medium, with a consistent and even pace... \textbf{B.} Medium, but with a noticeable deceleration... & 821 / 907 \\
RHY & \textbf{Q:} Analyze the rhythmic features of the speaker's speech. \textbf{A.} The rhythm is fluent and coherent, delivered as a seamless whole... \textbf{B.} The rhythm is fragmented and disjointed... & 999 / 979 \\
VT & \textbf{Q:} How would you describe the physical texture of the speaker's voice? \textbf{A.} Full and resonant, with a subtle, consistent grainy texture. \textbf{B.} Full and resonant, with an exceptionally clear... & 999 / 981 \\ \midrule

\multicolumn{3}{c}{\textbf{Affective and Semantic Reasoning}} \\ \midrule
EMO & \textbf{Q:} Based on the vocal performance and textual context, what is the main emotion conveyed? \textbf{A.} Defensive and exasperated \textbf{B.} Defensive yet deeply sorrowful... & 986 / 815 \\
TON & \textbf{Q:} What are the tonal characteristics of the voice in the speech? \textbf{A.} A bewildered inquiring tone \textbf{B.} A doubtful questioning tone... & 997 / 984 \\
CI & \textbf{Q:} Based on the vocal performance, what could be the context of this speech? \textbf{A.} A character is explaining the futility of escape... \textbf{B.} A character is lamenting a personal failure... & 999 / 996 \\ \midrule

\multicolumn{3}{c}{\textbf{Acoustic Scene Analysis}} \\ \midrule
BS & \textbf{Q:} Which of the following is the most accurate description of the speech's background sound? \textbf{A.} Light, playful background music \textbf{B.} Sounds of objects being moved and placed... & 523 / 503 \\ 
AE & \textbf{Q:} Based on the reverberation and sense of space, what is the most likely acoustic physical environment? \textbf{A.} Inferred as a studio or professional recording room... \textbf{B.} Inferred as a large, empty hall or church...  & 504 / 504 \\ \midrule

\multicolumn{3}{c}{\textbf{Linguistic-Paralinguistic Integration}} \\ \midrule
PE & \textbf{Q:} Which of the following paralinguistic events is present in the speech? \textbf{A.} The speaker makes a breathing sound. \textbf{B.} The speaker makes a sighing sound... & 976 / 778 \\
TPT & \textbf{Prompt:} Please transcribe the speech into text and provide tags for the paralinguistic events at their occurrence positions. \textbf{Output:} \textless Crying\textgreater~You gotta hide me. Death is after me.  & 976 / 778 \\ 
 \bottomrule
\end{tabularx}
\end{table*}

\subsection{FMSU-Bench}

To rigorously and systematically assess whether speech LLMs truly possess decoupled and complex auditory perception capabilities, we construct and release \textbf{FMSU-Bench}, a comprehensive, fine-grained, and multi-dimensional speech understanding benchmark. Comprising over 20,000 high-quality bilingual instances in Chinese and English, FMSU-Bench encompasses 14 independent dimensions of speech evaluation tasks structured into a 5-tier taxonomy, as detailed in Table~\ref{tab:fmsu_bench_tasks}. These 14 dimensions are abbreviated as follows: Gender (GEN), Age (AGE), Accent (ACC), Pitch (PIT), Speaking Rate (SR), Rhythm (RHY), Voice Texture (VT), Emotion (EMO), Tone (TON), Contextual Inference (CI), Background Sound (BS), Acoustic Environment (AE), Paralinguistic Events (PE), and Transcription with Paralinguistic Tags (TPT).

\subsubsection{Data Filtering and Correction}

To construct FMSU-Bench, we enforce clear inclusion criteria. From the corpus produced by our data curation pipeline, we retain only speech samples exhibiting a WER/CER below 10\% and a minimum duration of three seconds. Despite passing the aforementioned multi-expert cross-validation, these samples undergo a manual verification protocol. Two independent experts review each annotation, retaining the sample if both accept it unmodified. We discard samples if judgments diverge (i.e., only one expert proposes modifications). If both experts modify the annotation, a third senior expert verifies the semantic consistency of their revisions, retaining the sample only if the corrections align. Ultimately, this refinement process yields 500 to 1,000 manually verified test samples per attribute, establishing a highly reliable benchmark of over 20,000 bilingual instances.

\subsubsection{Task Formulation}
To ensure objective evaluation and circumvent LLM-as-a-Judge biases, we formulate 13 speech attribute tasks as MCQs, as illustrated in Table~\ref{tab:fmsu_bench_tasks}. To distinguish genuine auditory perception from text-based guessing, we leverage Gemini 2.5 Pro to synthesize diverse question stems paired with multiple fine-grained options. Every MCQ pairs the fine-grained ground truth with strategically designed distractors that target specific perceptual vulnerabilities: 
\begin{itemize}
    \item \textbf{Fine-Grained Acoustic Distractors:} These options intentionally conflate subtle acoustic nuances to rigorously challenge the model's micro-perception capabilities. For instance, they force the model to distinguish between closely related emotional undertones, such as confusing ``sad with a hint of repentance'' with ``sad with a hint of disappointment.''
    \item \textbf{Semantic Trap Distractors:} Fabricated entirely from textual semantics, these highly plausible choices deliberately disregard the actual acoustic signal. For example, if the textual transcript dictates, ``I am so happy today,'' but the utterance is delivered in a distinctly sorrowful tone, the semantic trap would deceptively label the emotion as ``joyful and full of passion.'' This mechanism precisely traps and penalizes models suffering from text-reliance hallucinations..
\end{itemize}

For the open-ended transcription task—Transcription with Paralinguistic Tags, we require the model to generate accurate textual transcriptions while simultaneously anchoring paralinguistic tags (e.g., \texttt{<Laughter>}, \texttt{<Crying>}) at their precise temporal locations,  as illustrated in Table~\ref{tab:fmsu_bench_tasks}. Notably, we deliberately incorporate control samples entirely devoid of paralinguistic events. This design serves as a robust probe to detect and penalize models prone to ``paralinguistic hallucinations" (i.e., fabricating non-existent paralinguistic tags).

\subsubsection{Evaluation Metrics}
For the 13 MCQ tasks listed in Table~\ref{tab:fmsu_bench_tasks}, we employ standard Accuracy as the evaluation metric. It is defined as the ratio of the number of correctly answered MCQs to the total number of MCQs per task, ensuring an objective and unbiased evaluation. However, for the Transcription with Paralinguistic Tags task, traditional WER/CER metrics fail to effectively measure the prediction and localization accuracy of paralinguistic events. To address this, we propose a novel composite evaluation metric: \textbf{Paralinguistic-Aware Transcription Accuracy ($PATA$)}. $PATA$ comprehensively evaluates performance by combining pure text transcription accuracy with paralinguistic tag prediction accuracy. Specifically, we first compute pure text transcription accuracy (removing all paralinguistic tags) as $\max(0, 1 - ERR_{text})$, where $ERR_{text}$ denotes the standard WER for English or CER for Chinese. To evaluate the paralinguistic events, we treat both text tokens (words/characters) and paralinguistic tags uniformly as individual sequence tokens. We then apply standard Levenshtein distance alignment between the generated and reference sequences, strictly retaining all paralinguistic tags within both sequences. A paralinguistic tag is recorded as a True Positive if and only if it correctly aligns with the reference tag in both relative sequential position and category. Based on this alignment, we compute the paralinguistic F1-score ($F1_{para}$). Finally, $PATA$ is formulated as a weighted linear combination of these two components:
\begin{equation}
    PATA = \alpha \cdot \max(0, 1 - ERR_{text}) + (1 - \alpha)  \cdot F1_{para}
\end{equation}
where we empirically set $\alpha = 0.5$ to place equal emphasis on semantic preservation and paralinguistic perception.

\subsection{FM-Speech}
Leveraging the high-quality, fine-grained corpus generated by our data curation pipeline, we introduce \textbf{FM-Speech}, built upon the frontier Qwen3-Omni~\cite{xu2025qwen3omni} architecture. Directly forcing a model to generate complex, multi-dimensional structured outputs during early training often triggers severe modality gaps and information overload. To mitigate this, this section details the \textbf{progressive curriculum fine-tuning framework} employed by FM-Speech. This framework systematically decouples complex auditory comprehension into three incremental training stages, driven by three distinct training data formulations.

\subsubsection{Training Data Formulation}
Based on the fine-grained multi-dimensional speech attribute JSON annotations produced by our data curation pipeline, we construct three distinct types of training data to progressively guide the model in building robust auditory perception capabilities:
\begin{itemize}
\item \textbf{Type I: Single-Dimension MCQs.} For each speech attribute (excluding transcription with paralinguistic tags), we design a MCQ with one fine-grained ground truth and multiple negatives that exhibit varying degrees of deviation from the ground truth (illustrative examples are detailed in the ``Example'' column of Table~\ref{tab:fmsu_bench_tasks}). This discriminative task enforces fundamental cross-modal alignment between low-level acoustic signals and specific textual concepts.
\item \textbf{Type II: Single-Dimension Open QA.} This requires the model to generate naturally phrased, detailed descriptions for an individual speech attribute (e.g., \textbf{Qusetion}: Analyze the rhythmic features of the speaker’s speech. \textbf{Answer}: The rhythm is fluent and coherent, delivered as a seamless whole...). It transitions the model from perceptual discrimination (MCQs) to independent generative articulation without candidate hints.
\item \textbf{Type III: Full-Dimensional JSON Generation.} Using 14-dimension fine-grained speech attribute annotations in a structured JSON format as targets, this task forces the model to concurrently process all decoupled acoustic features and adhere to strict formatting constraints.
\end{itemize}

\subsubsection{Dynamic Data-Mixing Strategy}
To mitigate the catastrophic forgetting induced by exposing the model to a single task formulation per stage during sequential training, we integrate a dynamic data-mixing strategy across all three training stages. This curriculum effectively balances fine-grained acoustic perception with holistic structural generation:
\begin{itemize}
\item \textbf{Stage 1: Warm-up.} Using 60\% Type I (MCQ) and 40\% Type II (Open QA) data, we temporarily exclude complex structured tasks. This stage forces the model to focus exclusively on acoustic-linguistic alignment for individual dimensions, establishing robust auditory intuition.
\item \textbf{Stage 2: Capability Ramp-up.} The distribution shifts to 20\% Type I, 40\% Type II, and 40\% Type III (JSON). By introducing full-dimensional JSONs while retaining single-dimension tasks, we guide the model to adapt to multi-feature integration and complex formatting without degrading its previously acquired fine-grained perception.
\item \textbf{Stage 3: Final Alignment.} The data distribution transitions to 100\% Type III (JSON). These multi-dimensional JSON annotations implicitly cover all the individual speech attributes introduced in prior stages. As the final alignment stage, the objective is to strictly lock in the output paradigm. This stage forces the model to seamlessly consolidate its fine-grained auditory perception into a holistic paradigm, enabling comprehensive analysis and properly formatted generation.
\end{itemize}
Through this progressive curriculum fine-tuning framework, FM-Speech successfully masters the highly challenging multi-dimensional JSON generation while preserving fine-grained auditory perception.

\section{Experimental Setup}

\subsection{Evaluation Setup for FMSU-Bench}
To comprehensively and objectively assess the true capabilities of diverse speech LLMs on fine-grained, multi-dimensional speech understanding tasks, we conduct large-scale systematic evaluations on FMSU-Bench. This section details the selected models, deployment environments, and adaptive evaluation protocols tailored to different task paradigms.

\subsubsection{Model Selection and Deployment Setup}
We comprehensively evaluate our proposed FM-Speech against 11 advanced speech LLMs, comprising eight mainstream open-source models (Audio Flamingo 3~\cite{goel2025audioflamingo3}, Qwen3-Omni~\cite{xu2025qwen3omni}, Kimi-Audio~\cite{ding2025kimi}, Step-Audio 2~\cite{wu2025stepaudio2}, Omni-Captioner~\cite{ma2025omnicaptioner}, Mimo-Audio~\cite{zhang2025mimo}, Qwen2.5-Omni~\cite{xu2025qwen25omnitechnicalreport}, and Qwen2-Audio~\cite{chu2024qwen2audio}) and three representative proprietary models (Gemini 2.5 Flash, Gemini 3 Flash, and Gemini 3.1 Pro~\cite{comanici2025gemini}). 
To ensure inference fairness and reproducibility, all open-source models (including FM-Speech) are deployed locally. We strictly adhere to their official configuration guidelines, hosting the inference services on compute nodes equipped with 8 NVIDIA L20 GPUs. Conversely, evaluations for the three proprietary models are conducted by rigorously invoking their official cloud APIs.

\subsubsection{Adaptive Evaluation Protocols}
Given the discrepancies in native output paradigms and instruction-following capabilities among the models, we design adaptive input and parsing protocols for the different task types in our FMSU-Bench. The \textbf{13 MCQ tasks} listed in Table~\ref{tab:fmsu_bench_tasks} objectively assess fine-grained, multi-dimensional speech understanding using strict accuracy metrics. For the vast majority of tested models, we directly input the question prompt, candidate options, and the speech signal, explicitly prompting the model to output the correct option. However, Omni-Captioner lacks the capacity to adapt to custom prompts, while our model is strictly aligned to output fine-grained, multi-dimensional structured JSONs. To accommodate these exceptions, we retain their native free-text responses generated from the test speech. Subsequently, we deploy Gemini 2.5 Pro as a response-to-option aligner to semantically map these free-text outputs to the most appropriate MCQ option. To ensure alignment accuracy, we conducted manual spot-checks on a random subset of 500 instances, yielding an agreement rate of 96\% between Gemini 2.5 Pro's alignments and human evaluations. 
For the \textbf{Transcription with Paralinguistic Tags task}, performance is measured via our proposed $PATA$ metric. For FM-Speech, we directly extract the transcription field from its JSON output. For other configurable models, we design explicit prompts guiding them to insert paralinguistic tags (e.g., \texttt{<Laughter>}, \texttt{<Crying>}) at precise temporal locations during transcription. Notably, Omni-Captioner is excluded from this evaluation, as its native paradigm neither supports custom prompting nor generates paralinguistic tags, rendering it architecturally incompatible with this task.

\subsection{Training Configuration of FM-Speech}

We initialize FM-Speech from Qwen3-Omni-30B-A3B-Instruct, a 30B-parameter multimodal LLM featuring a Mixture-of-Experts architecture that activates 3B parameters per forward pass~\cite{xu2025qwen3omni}. Efficient distributed training is conducted using the MS-Swift framework~\cite{reis2005swift} with a Megatron-LM backend across 8 NVIDIA L20 GPUs. Leveraging our robust data curation pipeline, we compile a large-scale corpus comprising approximately 2.3 million speech instances, each equipped with fine-grained, multi-dimensional speech attribute annotations. To facilitate the proposed progressive fine-tuning curriculum framework, we systematically restructure and augment these foundational annotations into three distinct training data paradigms: Type I (MCQs for Single Dimensions), Type II (Open-ended QA for Single Dimensions), and Type III ( Full-Dimensional JSON Generation).


We employ Low-Rank Adaptation (LoRA, $r=8, \alpha=32$) based on Qwen3-Omni-30B-A3B-Instruct through our three-stage curriculum training. In \textbf{Stage 1 Warm-up (3 Epochs)}, we establish foundational acoustic perception by keeping the LLM frozen and applying LoRA exclusively to the audio encoder and modality projector, utilizing 15M speech instances (9M Type I [60\%], 6M Type II [40\%]). For \textbf{Stages 2 \& 3 (6 Epochs each)}, we jointly apply LoRA across the LLM, audio encoder, and projector to achieve deep cross-modal fusion. Specifically, \textbf{Stage 2 Capability Ramp-up} serves as a transitional stage, leveraging 5.75M instances (1.15M Type I [20\%], 2.3M Type II [40\%], 2.3M Type III [40\%]) to smoothly introduce structured JSON generation. Subsequently, \textbf{Stage 3 Final Alignment} acts as the convergence stage, exclusively utilizing 2.3M Type III JSON instances to strictly lock in the target output paradigm. Across all training stages, we maintain a consistent global batch size of 128. The AdamW~\cite{loshchilov2017adamw} optimizer employs a cosine annealing decay schedule, setting the peak learning rate to $1 \times 10^{-5}$ and gradually decaying it to a minimum of $1 \times 10^{-6}$.

\begin{table*}[htbp]
\centering
\caption{Comprehensive evaluation results on the FMSU-Bench.}
\label{tab:main_results}
\setlength{\tabcolsep}{3.5pt}
\resizebox{\textwidth}{!}{
\begin{tabular}{@{} l | >{\columncolor{gray!15}}c | c | c c c | c c c c | c c c | c c | c c @{}}
\toprule
\textbf{Model} & \textbf{Avg (\%) $\uparrow$} & \textbf{Lang} & \textbf{GEN} & \textbf{AGE} & \textbf{ACC} & \textbf{PIT} & \textbf{SR} & \textbf{RHY} & \textbf{VT} & \textbf{EMO} & \textbf{TON} & \textbf{CI} & \textbf{BS} & \textbf{AE} & \textbf{PE} & \textbf{TPT} \\ \midrule

\multicolumn{17}{@{}l}{\textit{Proprietary Models}} \\ \midrule
\multirow{2}{*}{Gemini 3.1 Pro}   &  
& ZH & 86.3 & 62.4 & \underline{81.9} & 66.6 & 69.8 & \textbf{87.3} & 67.2 & \textbf{76.0} & \underline{65.6} & \textbf{93.1} & \textbf{66.7} & 75.6 & \underline{72.1} & \underline{71.0} \\
                                  & \multirow{-2}{*}{\textbf{74.0}} 
& EN & 85.3 & 65.2 & \textbf{75.8} & 74.6 & 71.7 & \textbf{91.9} & 76.4 & \underline{68.9} & 55.6 & \textbf{94.5} & \textbf{61.4} & \underline{85.1} & \underline{61.9} & 60.8 \\ \cmidrule{1-17}

\multirow{2}{*}{Gemini 3 Flash}   &  
& ZH & 85.4 & 63.3 & 65.1 & \textbf{69.3} & 74.7 & \underline{87.2} & \underline{71.9} & \underline{73.2} & 62.9 & \underline{86.9} & 60.2 & \textbf{76.6} & 63.3 & 69.5 \\
                                  & \multirow{-2}{*}{71.9} 
& EN & 88.3 & 59.1 & 67.0 & 77.7 & \textbf{75.5} & \underline{91.6} & \underline{80.2} & 61.3 & 56.6 & \underline{87.1} & 55.9 & \textbf{86.7} & 55.9 & \underline{61.4} \\ \cmidrule{1-17}

\multirow{2}{*}{Gemini 2.5 Flash} &  
& ZH & 97.0 & 53.4 & 55.4 & 62.1 & 68.1 & 80.3 & \textbf{75.8} & 69.6 & \textbf{77.8} & 85.5 & 46.4 & 70.2 & 53.9 & 63.5 \\
                                  & \multirow{-2}{*}{69.0} 
& EN & 93.6 & \underline{66.1} & 63.0 & \textbf{80.9} & 67.0 & 90.1 & \textbf{81.4} & 64.9 & \underline{58.2} & 84.9 & 51.1 & 66.7 & 52.1 & 52.4 \\ \midrule

\multicolumn{17}{@{}l}{\textit{Open-source Models}} \\ \midrule
\multirow{2}{*}{Audio Flamingo 3} &  
& ZH & 89.8 & 41.6 & 35.4 & 20.2 & 21.4 & 40.6 & 26.8 & 42.8 & 35.7 & 65.8 & 42.1 & 55.2 & 45.3 & 34.8 \\
                                  & \multirow{-2}{*}{47.6} 
& EN & 97.6 & 42.1 & 36.4 & 32.7 & 59.0 & 70.0 & 53.6 & 41.1 & 36.2 & 79.3 & 48.5 & 53.6 & 47.2 & 38.0 \\ \cmidrule{1-17}

\multirow{2}{*}{Kimi-Audio}       &  
& ZH & 79.4 & 42.4 & 59.3 & 37.2 & 37.9 & 49.9 & 23.9 & 49.7 & 45.1 & 56.0 & 50.1 & 71.0 & 59.9 & 56.8 \\
                                  & \multirow{-2}{*}{54.3} 
& EN & 95.6 & 50.1 & 54.4 & 49.7 & 51.9 & 74.5 & 38.9 & 51.9 & 42.6 & 69.6 & 58.5 & 67.9 & 53.0 & 44.2 \\ \cmidrule{1-17}

\multirow{2}{*}{Mimo-Audio}       &  
& ZH & 88.4 & 59.9 & 63.5 & 59.8 & 58.6 & 78.3 & 53.0 & 70.4 & 58.5 & 76.1 & 46.5 & \underline{75.8} & 50.6 & 50.8 \\
                                  & \multirow{-2}{*}{64.1} 
& EN & 96.9 & 57.1 & 64.8 & 64.6 & 65.9 & 81.5 & 60.2 & 64.3 & 54.6 & 76.6 & 50.3 & 79.2 & 41.7 & 47.4 \\ \cmidrule{1-17}

\multirow{2}{*}{Omni-Captioner}   &  
& ZH & 98.4 & 54.6 & 54.3 & 65.4 & 58.8 & 81.5 & 60.2 & 70.1 & 61.9 & 79.9 & 47.8 & 54.0 & 58.2 & -- \\
                                  & \multirow{-2}{*}{66.0} 
& EN & 98.8 & 48.6 & 65.0 & 71.1 & \underline{74.5} & 84.3 & 59.5 & \textbf{75.7} & 58.0 & 79.7 & 38.8 & 70.6 & 45.5 & -- \\ \cmidrule{1-17}

\multirow{2}{*}{Qwen2.5-Omni}     &  
& ZH & 96.9 & 61.0 & 70.9 & 63.8 & \underline{75.4} & 60.7 & 25.6 & 60.8 & 45.3 & 79.5 & 29.1 & 67.5 & 64.7 & 48.4 \\
                                  & \multirow{-2}{*}{59.7} 
& EN & \underline{99.3} & 62.4 & 47.9 & 45.5 & 70.6 & 70.5 & 39.0 & 48.7 & 40.7 & 87.0 & 48.7 & 66.7 & 49.7 & 45.0 \\ \cmidrule{1-17}

\multirow{2}{*}{Step-Audio 2}     &  
& ZH & 90.2 & 42.4 & 67.2 & 24.3 & 34.7 & 58.6 & 25.8 & 56.2 & 46.0 & 69.4 & 31.7 & 63.5 & 41.8 & 47.8 \\
                                  & \multirow{-2}{*}{48.7} 
& EN & 87.5 & 38.7 & 48.0 & 21.4 & 38.8 & 65.8 & 33.8 & 40.3 & 41.7 & 73.0 & 36.8 & 58.5 & 36.9 & 43.4 \\ \cmidrule{1-17}

\multirow{2}{*}{Qwen3-Omni}       &  
& ZH & \underline{99.0} & \underline{70.4} & 67.0 & \underline{68.2} & \textbf{75.6} & 76.0 & 58.0 & 70.6 & 57.3 & 80.1 & 57.6 & 72.2 & 65.2 & 61.1 \\
                                  & \multirow{-2}{*}{69.4} 
& EN & \textbf{99.5} & 64.8 & 54.3 & 73.0 & 74.2 & 84.4 & 70.4 & 59.4 & 57.6 & 80.2 & \underline{61.0} & 72.4 & 59.5 & 53.4 \\ \cmidrule{1-17}

\multirow{2}{*}{Qwen2-Audio}      &  
& ZH & 46.9 & 15.3 & 21.0 & 15.8 & 22.3 & 28.7 & 26.1 & 20.1 & 16.9 & 44.3 & 12.2 & 34.5 & 14.5 & 30.7 \\
                                  & \multirow{-2}{*}{23.6} 
& EN & 36.8 & 14.6 & 19.9 & 14.6 & 21.0 & 17.1 & 25.8 & 14.4 & 13.9 & 38.9 & 20.9 & 21.6 & 26.1 & 25.5 \\ \midrule

\multirow{2}{*}{\textbf{FM-Speech (Ours)}} &  
& ZH & \textbf{99.1} & \textbf{75.2} & \textbf{83.5} & 55.2 & 52.9 & 74.4 & 63.7 & 62.5 & 60.4 & 79.9 & \underline{65.8} & 70.8 & \textbf{87.5} & \textbf{77.0} \\
                                           & \multirow{-2}{*}{\underline{72.8}} 
& EN & \underline{99.3} & \textbf{79.2} & \underline{75.1} & \underline{78.4} & 72.2 & 78.6 & 63.6 & 63.3 & \textbf{65.4} & 72.5 & 55.8 & 77.1 & \textbf{69.9} & \textbf{79.8} \\ \bottomrule
\end{tabular}
}
\end{table*}

\section{Experimental Results}
\subsection{Results and Analysis on FMSU-Bench}
Table \ref{tab:main_results} presents a comprehensive performance comparison between our proposed FM-Speech and 11 frontier speech LLMs across 14 fine-grained tasks on both the Chinese (ZH) and English (EN) subsets of FMSU-Bench. For brevity, task abbreviations from Table \ref{tab:fmsu_bench_tasks} are adopted (e.g., TPT refers to the $PATA$ score for Transcription with Paralinguistic Tags, while GEN and ACC denote the MCQ accuracies for Gender and Accent, respectively). Within the table, the results for each model are organized into two rows: ZH (top) and EN (bottom). The best and second-best performing results are highlighted in \textbf{bold} and \underline{underlined}, respectively. Additionally, the \textbf{Avg} column reports the overall average score across all tasks and languages. A detailed analysis of these findings is provided in the subsequent subsections.

\textbf{Overall Difficulty and Discriminative Power.}
While the leading proprietary model, Gemini 3.1 Pro, achieves the highest overall performance across the benchmark, it still fails to fully resolve all 14 tasks. Meanwhile, several prominent open-source models (e.g., Kimi-Audio, Step-Audio 2, Audio Flamingo 3, Qwen2.5-Omni) struggle to surpass a 60\% average accuracy threshold, with the earlier Qwen2-Audio languishing at 23.6\%. This wide variance demonstrates that FMSU-Bench serves as a rigorous, discriminative touchstone for testing fine-grained, multi-dimensional speech understanding, highlighting substantial room for future algorithmic advancement.

\textbf{Performance Variation Across Dimensions.}
A detailed breakdown across the 14 speech attribute dimensions reveals a pronounced performance asymmetry. Most models excel in macro-statistical and semantic-dependent tasks. For instance, Qwen3-Omni achieves 99\% on the Chinese GEN task, and Gemini 3.1 Pro reaches 94.5\% on the English CI task. However, performance sharply collapses on tasks requiring micro-acoustic perception, acoustic scene analysis, and linguistic-paralinguistic integration (e.g., PIT, VT, BS, and TPT). These findings reveal a critical capability gap between semantic-level comprehension and true micro-acoustic perception. Overcoming this bottleneck requires shifting focus from text-dependent reasoning toward the deep disentanglement of nuanced acoustic and environmental features.

\textbf{Comparison of Open-Source and Proprietary Models.}
Notably, the performance gap between top-tier open-source and proprietary models is rapidly narrowing. Besides our proposed FM-Speech, the original open-source Qwen3-Omni secures a strong average score of 69.4\%, surpassing Gemini 2.5 Flash (69.0\%), and trailing only Gemini 3 Flash (71.9\%) and Gemini 3.1 Pro (74.0\%). This trend indicates that architectural barriers in fine-grained, multi-dimensional speech understanding are inherently surmountable. By leveraging highly efficient data curation and training strategies, open-source models are poised to further close this gap, potentially overtaking proprietary models with more efficient resource allocation.

\textbf{Effectiveness of the Proposed Method.}
Ultimately, the proposed FM-Speech achieves a remarkable average score of 72.8\% on FMSU-Bench, comprehensively outperforming all evaluated open-source models, 
Furthermore, it surpasses the proprietary Gemini 3 Flash (71.9\%) and closely approaches the performance of the industry-leading Gemini 3.1 Pro (74.0\%). Notably, FM-Speech secures either the SOTA or second-best position across multiple speech dimensions, such as GEN, AGE, ACC, TON, BS, PE, and TPT task. These results demonstrate that the data curated by our pipeline effectively empowers the model to navigate real-world acoustic complexities. Furthermore, the progressive curriculum fine-tuning transitions the model from shallow text-dependence to deep acoustic decoupling, enhancing fine-grained, multi-dimensional speech comprehension.

\begin{table}[htbp]
\centering
\caption{Ablation study results on FMSU-Bench.}
\label{tab:ablation}
\small
\begin{tabular}{@{}lcc@{}}
\toprule
\textbf{Model Configuration} & \textbf{Params} & \textbf{Avg (\%) $\uparrow$} \\ \midrule
Original Qwen3-Omni          & 30B-A3B             & 69.4            \\
Fine-tuned Qwen3-Omni (Single Stage)    & 30B-A3B             & 67.8            \\
\textbf{FM-Speech (Full Three-Stage)} & 30B-A3B    & \textbf{72.8}   \\ \midrule
Original Qwen2.5-Omni        & 7B              & 59.7            \\
Fine-tuned Qwen2.5-Omni (Single Stage)    & 7B             & 55.2            \\
Fine-tuned Qwen2.5-Omni (Full Three-Stage) & 7B              & 63.9   \\ \bottomrule
\end{tabular}
\end{table}

\subsection{Ablation Study}

To evaluate the effectiveness and broad applicability of the Progressive Curriculum Fine-Tuning framework employed in FM-Speech, we conduct ablation studies on FMSU-Bench. Table \ref{tab:ablation} summarizes the overall experimental design and results.

We evaluate the necessity of our progressive three-stage curriculum fine-tuning by comparing three configurations: the original Qwen3-Omni (69.4\%), a variant fine-tuned exclusively on Type III full-dimensional JSON data (Single Stage), and our complete pipeline (FM-Speech). The ``Single Stage'' variant yields only 67.8\%. Lacking prior modality alignment, the model suffers from information overload and text-reliance hallucinations when navigating complex structural constraints. In contrast, our full three-stage model achieves 72.8\%. By gradually transitioning from single-attribute perception to multi-dimensional generation, the model effectively bridges the modality gap, demonstrating the indispensability of  progressive curriculum fine-tuning.

To validate the architecture-agnostic nature of our training framework, we apply the identical pipeline to Qwen2.5-Omni. Notably, this transitions the baseline from a 30B MoE architecture to a 7B dense architecture. While the original Qwen2.5-Omni baseline achieves 59.7\%, our full three-stage framework significantly elevates its performance to 63.9\%. Conversely, bypassing the curriculum (i.e., Single Stage) causes performance to plummet to 55.2\%, confirming that without prior modality alignment, even smaller dense models suffer severely from information overload. These results robustly validate the scalability and cross-architecture generalization of our progressive curriculum fine-tuning framework.

\section{conclusion}
This paper presents a holistic framework for fine-grained, real-world speech understanding. First, we develop a robust curation pipeline to extract structurally annotated speech from noisy audiovisuals. Second, we construct \textbf{FMSU-Bench}, a pioneering 14-dimension benchmark to rigorously evaluate speech LLMs. Third, we introduce \textbf{FM-Speech}, which employs a progressive fine-tuning framework on our curated corpus to resolve cross-modal overload and mitigate text-reliance hallucinations. Evaluations demonstrate that FM-Speech substantially outperforms open-source models. By bridging gaps across data, evaluation, and modeling, our work lays a robust foundation for future highly perceptive speech models.

\bibliographystyle{splncs04}
\bibliography{ref.bib}

@inproceedings{panayotov2015librispeech,
  title={Librispeech: an asr corpus based on public domain audio books},
  author={Panayotov, Vassil and Chen, Guoguo and Povey, Daniel and Khudanpur, Sanjeev},
  booktitle={2015 IEEE international conference on acoustics, speech and signal processing (ICASSP)},
  pages={5206--5210},
  year={2015},
  organization={IEEE}
}

@inproceedings{he2024emilia,
  title={Emilia: An extensive, multilingual, and diverse speech dataset for large-scale speech generation},
  author={He, Haorui and Shang, Zengqiang and Wang, Chaoren and Li, Xuyuan and Gu, Yicheng and Hua, Hua and Liu, Liwei and Yang, Chen and Li, Jiaqi and Shi, Peiyang and others},
  booktitle={2024 IEEE Spoken Language Technology Workshop (SLT)},
  pages={885--890},
  year={2024},
  organization={IEEE}
}

@article{wu2025stepaudio2,
  title={Step-audio 2 technical report},
  author={Wu, Boyong and Yan, Chao and Hu, Chen and Yi, Cheng and Feng, Chengli and Tian, Fei and Shen, Feiyu and Yu, Gang and Zhang, Haoyang and Li, Jingbei and others},
  journal={arXiv preprint arXiv:2507.16632},
  year={2025}
}

@article{ding2025kimi,
  title={Kimi-audio technical report},
  author={Ding, Ding and Ju, Zeqian and Leng, Yichong and Liu, Songxiang and Liu, Tong and Shang, Zeyu and Shen, Kai and Song, Wei and Tan, Xu and Tang, Heyi and others},
  journal={arXiv preprint arXiv:2504.18425},
  year={2025}
}

@article{zhang2025mimo,
  title={MiMo-Audio: Audio Language Models are Few-Shot Learners},
  author={Zhang, Dong and Wang, Gang and Xue, Jinlong and Fang, Kai and Zhao, Liang and Ma, Rui and Ren, Shuhuai and Liu, Shuo and Guo, Tao and Zhuang, Weiji and others},
  journal={arXiv preprint arXiv:2512.23808},
  year={2025}
}

@inproceedings{yang2024air,
  title={Air-bench: Benchmarking large audio-language models via generative comprehension},
  author={Yang, Qian and Xu, Jin and Liu, Wenrui and Chu, Yunfei and Jiang, Ziyue and Zhou, Xiaohuan and Leng, Yichong and Lv, Yuanjun and Zhao, Zhou and Zhou, Chang and others},
  booktitle={Proceedings of the 62nd Annual Meeting of the Association for Computational Linguistics (Volume 1: Long Papers)},
  pages={1979--1998},
  year={2024}
}

@article{sakshi2024mmau,
  title={Mmau: A massive multi-task audio understanding and reasoning benchmark},
  author={Sakshi, Sakshi and Tyagi, Utkarsh and Kumar, Sonal and Seth, Ashish and Selvakumar, Ramaneswaran and Nieto, Oriol and Duraiswami, Ramani and Ghosh, Sreyan and Manocha, Dinesh},
  journal={arXiv preprint arXiv:2410.19168},
  year={2024}
}

@article{ma2025mmar,
  title={Mmar: A challenging benchmark for deep reasoning in speech, audio, music, and their mix},
  author={Ma, Ziyang and Ma, Yinghao and Zhu, Yanqiao and Yang, Chen and Chao, Yi-Wen and Xu, Ruiyang and Chen, Wenxi and Chen, Yuanzhe and Chen, Zhuo and Cong, Jian and others},
  journal={arXiv preprint arXiv:2505.13032},
  year={2025}
}

@article{liao2025nvspeech,
  title={Nvspeech: An integrated and scalable pipeline for human-like speech modeling with paralinguistic vocalizations},
  author={Liao, Huan and Ni, Qinke and Wang, Yuancheng and Lu, Yiheng and Zhan, Haoyue and Xie, Pengyuan and Zhang, Qiang and Wu, Zhizheng},
  journal={arXiv preprint arXiv:2508.04195},
  year={2025}
}

@inproceedings{wu2025smiip,
  title={SMIIP-NV: A Multi-Annotation Non-Verbal Expressive Speech Corpus in Mandarin for LLM-Based Speech Synthesis},
  author={Wu, Zhuojun and Liu, Dong and Liu, Juan and Wang, Yechen and Li, Linxi and Jin, Liwei and Bu, Hui and Zhang, Pengyuan and Li, Ming},
  booktitle={Proceedings of the 33rd ACM International Conference on Multimedia},
  pages={12564--12570},
  year={2025}
}

@article{ye2025scalable38k,
  title={A scalable pipeline for enabling non-verbal speech generation and understanding},
  author={Ye, Runchuan and Zhou, Yixuan and Yu, Renjie and Lin, Zijian and Li, Kehan and Li, Xiang and Liu, Xin and Zeng, Guoyang and Wu, Zhiyong},
  journal={arXiv preprint arXiv:2508.05385},
  year={2025}
}

@article{borisov2025nonverbaltts,
  title={Nonverbaltts: A public english corpus of text-aligned nonverbal vocalizations with emotion annotations for text-to-speech},
  author={Borisov, Maksim and Spirin, Egor and Diatlova, Daria},
  journal={arXiv preprint arXiv:2507.13155},
  year={2025}
}

@inproceedings{ardila2020commonvoice,
  title={Common voice: A massively-multilingual speech corpus},
  author={Ardila, Rosana and Branson, Megan and Davis, Kelly and Kohler, Michael and Meyer, Josh and Henretty, Michael and Morais, Reuben and Saunders, Lindsay and Tyers, Francis and Weber, Gregor},
  booktitle={Proceedings of the twelfth language resources and evaluation conference},
  pages={4218--4222},
  year={2020}
}

@article{comanici2025gemini,
  title={Gemini 2.5: Pushing the frontier with advanced reasoning, multimodality, long context, and next generation agentic capabilities},
  author={Comanici, Gheorghe and Bieber, Eric and Schaekermann, Mike and Pasupat, Ice and Sachdeva, Noveen and Dhillon, Inderjit and Blistein, Marcel and Ram, Ori and Zhang, Dan and Rosen, Evan and others},
  journal={arXiv preprint arXiv:2507.06261},
  year={2025}
}

@misc{Silero-VAD,
  author = {Silero Team},
  title = {Silero VAD: pre-trained enterprise-grade Voice Activity Detector (VAD), Number Detector and Language Classifier},
  year = {2024},
  publisher = {GitHub},
  journal = {GitHub repository},
  howpublished = {\url{https://github.com/snakers4/silero-vad}},
  commit = {insert_some_commit_here},
  email = {hello@silero.ai}
}

@inproceedings{li2026wenetspeechyue,
  title={Wenetspeech-yue: A large-scale cantonese speech corpus with multi-dimensional annotation},
  author={Li, Longhao and Guo, Zhao and Chen, Hongjie and Dai, Yuhang and Zhang, Ziyu and Xue, Hongfei and Zuo, Tianlun and Wang, Chengyou and Wang, Shuiyuan and Xu, Xin and others},
  booktitle={Proceedings of the AAAI Conference on Artificial Intelligence},
  volume={40},
  number={37},
  pages={31627--31635},
  year={2026}
}

@article{dai2025wenetspeechchuan,
  title={Wenetspeech-chuan: A large-scale sichuanese corpus with rich annotation for dialectal speech processing},
  author={Dai, Yuhang and Zhang, Ziyu and Wang, Shuai and Li, Longhao and Guo, Zhao and Zuo, Tianlun and Wang, Shuiyuan and Xue, Hongfei and Wang, Chengyou and Wang, Qing and others},
  journal={arXiv preprint arXiv:2509.18004},
  year={2025}
}

@article{wang2026wenetspeechwu,
  title={WenetSpeech-Wu: Datasets, Benchmarks, and Models for a Unified Chinese Wu Dialect Speech Processing Ecosystem},
  author={Wang, Chengyou and Shao, Mingchen and Hu, Jingbin and Zhu, Zeyu and Xue, Hongfei and Mu, Bingshen and Xu, Xin and Duan, Xingyi and Zhang, Binbin and Zhu, Pengcheng and others},
  journal={arXiv preprint arXiv:2601.11027},
  year={2026}
}

@article{shi2026qwen3asr,
  title={Qwen3-ASR Technical Report},
  author={Shi, Xian and Wang, Xiong and Guo, Zhifang and Wang, Yongqi and Zhang, Pei and Zhang, Xinyu and Guo, Zishan and Hao, Hongkun and Xi, Yu and Yang, Baosong and others},
  journal={arXiv preprint arXiv:2601.21337},
  year={2026}
}

@inproceedings{ma2024emotion2vec,
  title={emotion2vec: Self-supervised pre-training for speech emotion representation},
  author={Ma, Ziyang and Zheng, Zhisheng and Ye, Jiaxin and Li, Jinchao and Gao, Zhifu and Zhang, Shiliang and Chen, Xie},
  booktitle={Findings of the Association for Computational Linguistics: ACL 2024},
  pages={15747--15760},
  year={2024}
}

@article{tian2025stepaudior1,
  title={Step-Audio-R1 Technical Report},
  author={Tian, Fei and Zhang, Xiangyu Tony and Zhang, Yuxin and Zhang, Haoyang and Li, Yuxin and Liu, Daijiao and Deng, Yayue and Wu, Donghang and Chen, Jun and Zhao, Liang and others},
  journal={arXiv preprint arXiv:2511.15848},
  year={2025}
}

@article{feng2025voxprofile,
  title={Vox-profile: A speech foundation model benchmark for characterizing diverse speaker and speech traits},
  author={Feng, Tiantian and Lee, Jihwan and Xu, Anfeng and Lee, Yoonjeong and Lertpetchpun, Thanathai and Shi, Xuan and Wang, Helin and Thebaud, Thomas and Moro-Velazquez, Laureano and Byrd, Dani and others},
  journal={arXiv preprint arXiv:2505.14648},
  year={2025}
}

@article{xu2025qwen3omni,
  title={Qwen3-omni technical report},
  author={Xu, Jin and Guo, Zhifang and Hu, Hangrui and Chu, Yunfei and Wang, Xiong and He, Jinzheng and Wang, Yuxuan and Shi, Xian and He, Ting and Zhu, Xinfa and others},
  journal={arXiv preprint arXiv:2509.17765},
  year={2025}
}

@inproceedings{reis2005swift,
  title={SWIFT: Software implemented fault tolerance},
  author={Reis, George A and Chang, Jonathan and Vachharajani, Neil and Rangan, Ram and August, David I},
  booktitle={International symposium on Code generation and optimization},
  pages={243--254},
  year={2005},
  organization={IEEE}
}

@article{loshchilov2017adamw,
  title={Decoupled weight decay regularization},
  author={Loshchilov, Ilya and Hutter, Frank},
  journal={arXiv preprint arXiv:1711.05101},
  year={2017}
}

@article{goel2025audioflamingo3,
  title={Audio flamingo 3: Advancing audio intelligence with fully open large audio language models},
  author={Goel, Arushi and Ghosh, Sreyan and Kim, Jaehyeon and Kumar, Sonal and Kong, Zhifeng and Lee, Sang-gil and Yang, Chao-Han Huck and Duraiswami, Ramani and Manocha, Dinesh and Valle, Rafael and others},
  journal={arXiv preprint arXiv:2507.08128},
  year={2025}
}

@article{ma2025omnicaptioner,
  title={Omni-Captioner: Data Pipeline, Models, and Benchmark for Omni Detailed Perception},
  author={Ma, Ziyang and Xu, Ruiyang and Xing, Zhenghao and Chu, Yunfei and Wang, Yuxuan and He, Jinzheng and Xu, Jin and Heng, Pheng-Ann and Yu, Kai and Lin, Junyang and others},
  journal={arXiv preprint arXiv:2510.12720},
  year={2025}
}

@misc{xu2025qwen25omnitechnicalreport,
      title={Qwen2.5-Omni Technical Report}, 
      author={Jin Xu and Zhifang Guo and Jinzheng He and Hangrui Hu and Ting He and Shuai Bai and Keqin Chen and Jialin Wang and Yang Fan and Kai Dang and Bin Zhang and Xiong Wang and Yunfei Chu and Junyang Lin},
      url={https://arxiv.org/abs/2503.20215}, 
      year={2025}
}

@article{chu2024qwen2audio,
  title={Qwen2-audio technical report},
  author={Chu, Yunfei and Xu, Jin and Yang, Qian and Wei, Haojie and Wei, Xipin and Guo, Zhifang and Leng, Yichong and Lv, Yuanjun and He, Jinzheng and Lin, Junyang and others},
  journal={arXiv preprint arXiv:2407.10759},
  year={2024}
}

\end{document}